\documentclass[12pt]{article}
\usepackage{amscd,amssymb,amsmath,latexsym,enumerate}
\usepackage[mathscr]{euscript}
\usepackage{epsfig}
\usepackage{fancybox}
\usepackage{verbatim}

\usepackage[normalem]{ulem}
\usepackage{multirow}
\usepackage{stackengine}

\usepackage[small,nohug]{diagrams}
\diagramstyle[labelstyle=\scriptstyle]

\usepackage{color}
\newcommand{\red}{\textcolor[rgb]{0.99,0.00,0.00}}

\usepackage{xcolor}
\definecolor{MyBlue}{cmyk}{1,0.13,0,0.63}
\definecolor{MyGreen}{cmyk}{0.91,0,0.88,0.52}
\newcommand{\mylinkcolor}{MyBlue}
\newcommand{\mycitecolor}{MyGreen}
\newcommand{\myurlcolor}{black}

\usepackage{hyperref}
\hypersetup{%
  bookmarksnumbered=true,bookmarksopen=false,%
  plainpages=false,
  linktocpage=true,%
  colorlinks=true,breaklinks=true,%
  linkcolor=\mylinkcolor,citecolor=\mycitecolor,urlcolor=\myurlcolor,%
  pdfpagelayout=OneColumn,%
  pageanchor=true,%
}

\newcommand{\Exp}{{\mbox{\rm Exp}}}

\title{The generators of the $K$-groups of the sphere}

\author{Hermann Schulz-Baldes$^1$ and Tom Stoiber$^2$
\\
{\small $^1$Department Mathematik, Friedrich-Alexander-Universit\"at Erlangen-N\"urnberg, Germany}
\\
{\small Email: schuba@mi.uni-erlangen.de}
\\
{\small $^2$Department of Physics, Yeshiva Universtity, New York, USA}
}

\date{ }

\newtheorem{proposi}{Proposition}
\newtheorem{lemma}{Lemma}

\newcommand{\BM}{{\mathbb B}}
\newcommand{\CM}{{\mathbb C}}

\newcommand{\RM}{{\mathbb R}}
\newcommand{\SM}{{\mathbb S}}
\newcommand{\TM}{{\mathbb T}}
\newcommand{\ZM}{{\mathbb Z}}

\newcommand{\KM}{{\mathbb K}}

\newcommand{\DM}{{\mathbb D}}

\newcommand{\Aa}{{\cal A}}

\newcommand{\Vv}{{\cal V}}

\newcommand{\Ii}{{\cal I}}

\newcommand{\Qq}{{\cal Q}}
\newcommand{\Kk}{{\cal K}}
\newcommand{\Hh}{{\cal H}}

\newcommand{\one}{{\bf 1}}

\newcommand{\Ch}{{\rm Ch}} 
\newcommand{\Ind}{{\rm Ind}}

\newcommand{\diag}{{\rm diag}}

\newcommand{\Alge}{{\cal B}}

\begin{document}

\maketitle

\begin{abstract}
This note presents an elementary iterative construction of the generators for the complex $K$-groups $K_i(C(\SM^d))$ of the $d$-dimensional spheres. These generators are explicitly given as the restrictions of Dirac or Weyl Hamiltonians to the unit sphere. Connections to solid state physics are briefly elaborated on. 
\end{abstract}

\vspace{.5cm}

\section{Introduction}

Recall \cite{WO,Par} that the complex $K$-groups on the sphere $\SM^d=\{x\in\RM^{d+1}:\|x\|=1\}$ are given by
\begin{equation}
	\label{eq-KSphere}
	K_0(C(\SM^d))
	\;=\;
	\left\{
	\begin{array}{cc}
		0\;, & d \mbox{ odd}\;,
		\\
		\ZM\oplus\ZM\;,  & d \mbox{ even}\;,
	\end{array}
	\right.
	\qquad
	K_1(C(\SM^d))
	\;=\;
	\left\{
	\begin{array}{cc}
		\ZM\;, & d \mbox{ odd}\;,
		\\
		0\;,  & d \mbox{ even}\;,
	\end{array}
	\right.
\end{equation}
where $C(\SM^d)$ denotes the set of continuous functions on $\SM^d$. For even $d$, one of the summands in $K_0(C(\SM^d))$ is generated by the identity and hence {is} trivial. By definition, the reduced $K$-group $\widetilde{K}_0(C(\SM^d))\cong\ZM$ only contains the non-trivial summand. Surprisingly, we could not find a list of an explicit expression of the generators of $K_i(C(\SM^d))$ in the literature. In fact, in the standard textbooks \cite{Bla, WO, RLL} on operator $K$-theory the computation has been delegated to exercises or {is only given for} low-dimensional cases. {This is unfortunate as}  the computations are prone to sign ambiguities {which require to} carefully fix certain conventions.  {On the other hand,} the generators are often useful for practical computations ({\it e.g.} in connection with monopoles \cite{CSB} or for a proof of the Callias index theorem which by a homotopy argument can essentially be reduced to an explicit verification for the generators \cite{SS1}).  This  note aims to explicitly derive these generators using only the definition of the $K$-groups and the connecting maps at a level that is accessible to a reader who is familiar with the introductory chapters of either of the above textbooks. The idea of the iterative construction in the dimension $d$ is the same as for the construction of the generators of the $K$-groups on the $d$-dimensional tori and rotation algebras, as spelled out in \cite{PS}.

\vspace{.2cm}

Let us recall two methods used to compute generators of the $K$-groups of the spheres:
\begin{enumerate}

\item[(i)] The stereographic projection $\RM^d\to \SM^d\setminus \{e_{d+1}\}$ identifies $\SM^d$ with the one-point compactification of $\RM^d$ and thus $C(\SM^d)$ with the unitization of $C_0(\RM^d)$. Therefore 
$$
K_*(C(\SM^d))\,\cong\, K_*(C_0(\RM^d)) \oplus K_*(\CM)
\;.
$$
From the elementary standpoint taken here, this merely states that generators for the $K$-groups of $\RM^d$ and $\SM^d$ are equally difficult to compute. Standard text books \cite{Bla,RLL,WO} compute the $K$-groups by iterating the Bott map respectively the suspension map $K_i(C_0(\RM^d))\to K_{i-1}(C_0(\RM^{d+1}))$, without providing explicit formulas for the generators. Roughly stated, this is what is carried out on the following few pages.

\item[(ii)] For $i=d\,\mbox{\rm mod}\, 2$ one can define a Chern homomorphism $\Ch: K_i(C(\SM^d)) \to \CM$ \cite{Get,K1,BvE}   which restricts to an isomorphism $\Ch: \widetilde{K}_i(C(\SM^d)) \to \ZM$ of reduced $K$-theory when properly normalized. Therefore one can find generators by guessing them and verifying that their Chern numbers are $\pm 1$, but this reasoning and computations are fairly intricate. 

\item[(iii)] Another way to obtain the generators is to resort to $KK$-theory which is known to provide representations of $K$-groups as well.  One can  then compute iteratively the Kasparov products $KK_i(\CM,C_0(\RM^d))\otimes KK_1(\CM,C_0(\RM)) \to KK_{1-i}(\CM,C_0(\RM^d)\otimes C_0(\RM))$, provided one knows the (possibly unbounded) representative of $KK_1(\CM,C_0(\RM))$. Even though certainly known to experts, this argument is also far from elementary. Here in Section~\ref{sec:fredholm}, the reasoning is reversed and the generators on the spheres are used to determine those of $KK_i(\CM,C_0(\RM^d))$.

\end{enumerate}

The aim is thus to show that the Dirac and Weyl Hamiltonians in dimension $d+1$, when restricted to the sphere, are generators of the two non-trivial summands $\ZM$ in \eqref{eq-KSphere}. While this is a well-known fact in the community, the iterative construction provided below does not seem to be spelled out anywhere.

\begin{proposi}
\label{prop-Generators}
For $d\geq 2$ even, let $\Gamma_1,\ldots,\Gamma_{d+1}$ be a selfadjoint irreducible representation of the Clifford algebra $\CM_{d+1}$. A generator of $\widetilde{K}_0(C(\SM^d))$ is given by the selfadjoint unitary 
$$
x\,\in\,\SM^{d}\,\mapsto\,Q_d^W(x)\,=\,\sum_{j=1}^{d+1}x_j\,\Gamma_j
\qquad
\mbox{\rm (Weyl Hamiltonian)}
\;.
$$
For odd $d$, let $\Gamma_1,\ldots,\Gamma_{d}$ be a selfadjoint irreducible representation of the Clifford algebra $\CM_{d}$, a generator of ${K}_1(C(\SM^d))$ is given by the unitary 
$$
x\,\in\,\SM^{d}\,\mapsto\,U_d^D(x)\,=\,\sum_{j=1}^{d}x_j\,\Gamma_j\,+\,\imath\, x_{d+1} \one
\qquad
\mbox{\rm (Dirac phase)}
\;.
$$
\end{proposi}

Let us note that odd $K$-groups can also be described by chiral selfadjoint unitaries. In this representation and for odd $d$, the above generator is  given by $x\in\SM^{d}\mapsto Q_d^D(x)=\sum_{j=1}^{d+1}x_j\,\Gamma_j$ which indeed anti-commutes with the self-adjoint unitary $\Gamma_{d+2}=\imath^{\frac{d+2}{2}}\Gamma_1 \cdots \Gamma_{d+1}$. Hence in the grading of $\Gamma_{d+2}$, one has $Q_d^D=\binom{\;0 \;\;(U^D_d)^*}{U^D_d\;\;\;\;0\;\;}$ which is hence the Dirac Hamiltonian. Let us also remark that the equivalent of Proposition~\ref{prop-Generators} for the van Daele $K$-groups of "real" spheres was recently given in \cite{JM} using very different methods. More precisely, the generators can be given in an analogous form if certain reality conditions on the Clifford representation are enforced. Presumably this can also be derived using the methods of this article by simply implementing the symmetries in the connecting maps as in \cite{BL,GS}.


\vspace{.2cm}

The proof that the elements given in Proposition~\ref{prop-Generators} are generators will  be carried out inductively in the dimension $d$, starting with $d=1$ for which it is known that ${K}_1(C(\SM^1))$ is generated by $x\in\SM^1\mapsto U_1^D(x)=x_1+\imath x_2$. The iteration uses the connecting maps of the short exact sequence  of C$^*$-algebras
\begin{equation}
\label{eq-SphereExact}
0 \;\rightarrow \;C_0(\DM^{d+1}) \;\overset{i}{\rightarrow}\; C(\overline{\DM^{d+1}})\;\overset{\pi}{\rightarrow} \;C(\SM^d) \;\rightarrow \;0 
\end{equation}
Here the $d$-dimensional open and closed unit discs (w.r.t. the euclidean norm $\|\,.\,\|$ as above) are given by
$$
\DM^d\;=\;
\{x\in\RM^d\;:\;\|x\|<1\}
\;,
\qquad
\overline{\DM^d}\;=\;
\{x\in\RM^d\;:\;\|x\|\leq 1\}
\;.
$$
The exact sequence simply reflects that one has the disjoint decomposition $\overline{\DM^{d+1}}={\DM^{d+1}}\cup\SM^d$. For the iterative construction,  convenient forms of the connecting maps will be spelled out in Section~\ref{sec:connectingmap}.

\vspace{.2cm}

In non-commutative geometry, the generators of Proposition~\ref{prop-Generators} are the restrictions to the unit sphere of the so-called dual Dirac or Clifford operators which represent the generator of $K_i(\RM^d)$ in the unbounded Fredholm picture of $K$-theory, {\it i.e.} under the isomorphism $K_i(\RM^d)\cong KK_i(\CM, \RM^d)$. The connection of those with the generators of the sphere will be made more explicit in Section~\ref{sec:fredholm}. On the other hand, in solid state physics the generators are rather viewed as the restrictions to the unit sphere of the Fourier transforms of the Weyl and Dirac Hamiltonians on $\RM^d$ which provide a good local description of so-called Weyl points and Dirac points. In a solid state framework, a $K$-theoretic description of these points and an analysis of their stability apparently goes back to \cite{Horava}. The importance of such points as topological phase transition points through semimetals has been stressed in many works (for a mathematical treatment, see \cite{PS,SS1}). In the remainder of the introduction, let us describe the solid state perspective on the generators in the physically most relevant dimensions $d=3$ and $d=2$. Let us first consider the three-dimensional so-called Weyl Hamiltonian 
$$
H\; =\; 
\sum_{j=1}^3  {p_j\otimes \sigma_j}
$$
where $\sigma_j$ are the three Pauli matrices  and the momentum operators are  $p_j = -\imath \nabla_j$. Densely defined on a suitable Sobolev space, $H$ is a self-adjoint operator on $L^2(\RM^3)\otimes \CM^2$. It models a three-dimensional massless particle, a so-called Weyl fermion, and is popular as an effective model for certain quasi-particles in solid state physics. It is convenient to consider it in the Fourier-transformed picture and therefore instead interprete $p_j$ as the (unbounded) multiplication operator by the $j$-th coordinate on $L^2(\RM^d)\otimes \CM^2$. In that way, $H$ is multiplication by the matrix-valued function 
$$
h: \RM^3 \to M_2(\CM)\;,
\qquad h(x) \;=\; \sum_{j=1}^3  \sigma_j x_j \;=\; 
\begin{pmatrix}
x_3 & x_1 + \imath x_2 \\
x_1 - \imath x_2 & -x_3
\end{pmatrix}
\;.
$$
The pointwise spectrum is $\sigma(h(x))=\{- \sqrt{\lVert x\rVert},\sqrt{\lVert x\rVert}\}$ and therefore there are two distinct sheets that meet at the point $x=0$. As is well-known in physics this band-crossing is stable: continuously modifying the function $h$ in a neighborhood of $x=0$ while keeping the asymptotic behavior fixed, the two sheets will always meet in at least one point. The reason for that is that the band-crossing carries a "topological charge" given by the Chern number which prevents the opening of a gap in the spectrum. To understand this in terms of $K$-theory let us note that $h\rvert_{\SM^2}$ is a self-adjoint unitary and therefore defines a $K$-theory class $[h\rvert_{\SM^2}]_0 \in K_0(C(\SM^2))$. The exact sequence \eqref{eq-SphereExact} for $d=2$ links the interior of the sphere (the open disk $\DM^3$) with its boundary $\partial\DM^3=\SM^2$. From the associated $K$-theoretic six-term exact sequence one can read off (as we will do in Section~\ref{sec:sixterm}) that the connecting homomorphism $\Exp: K_0(C(\SM^2))\to K_1(C_0(\DM^3))$ is a surjection which maps a generator of the reduced $K$-group $\widetilde{K}_0(C(\SM^2))$ to a generator of $K_1(C_0(\DM^3))$. Now recall that the computation of $\Exp$ requires one to continuously extend $h\rvert_{\SM^2}$ to the interior of the disk and if the image under $\Exp$ is supposed to be non-trivial, then no such extension can have a spectral gap inside the interval $[-1,1]$. Therefore the phenomenology is consistent with the fact that $[h\rvert_{\SM^2}]_0$ is a non-trivial vector bundle on the sphere (and indeed this is the generator from Proposition~\ref{prop-Generators}).

\vspace{.2cm}

Due to $\widetilde{K}_0(C(\SM^2))\cong \ZM$, this procedure associates an integer-valued "charge" to the band-crossing. In solid state physics a general three-dimensional translation-invariant Hamiltonian corresponds in momentum space to a function $h: X\to M_N(\CM)$ where $X$ is either $\RM^3$ or $\TM^3$ and the spectrum as a function of $x\in X$ could have multiple band-crossings  $x_1,\ldots ,x_n \in X$ at spectral values $e_1,\ldots ,e_n \in \RM$. To each isolated band-crossing $x_i$, one can associate a charge in $K_0(\SM^2)$. If one deforms $h$ continuously in such a way that all band-crossings stay isolated ({\it i.e.} zero-dimensional), then it is only possible to eliminate non-trivial band-crossings and separate the corresponding bands by merging two or more of them in such a way that their charges add up to $0$. In this sense, Weyl points, {\it i.e.} band-crossings around which a Hamiltonian locally can be approximated by a Weyl Hamiltonian, are stable features of band structures.

\vspace{.2cm}

In dimension $d=2$ one can similarly construct an element of $K_1(C(\SM^{1}))$ from the Dirac Hamiltonian
$H = \sum_{j=1}^2 \sigma_j p_j$ on $L^2(\RM^2)\otimes \CM^2$, which upon Fourier transform is again a multiplication operator by the matrix function 
$$
h: \RM^2 \to M_2(\CM)\;,
\qquad h(x) \;=\; \sum_{j=1}^2 \sigma_j x_j \;=\;
 \begin{pmatrix}
	0 & x_1 + \imath x_2 \\
	x_1 - \imath x_2 & 0
\end{pmatrix}
\;.
$$
As above, the spectrum consists of two sheets that meet at $x=0$. Such band-crossings that locally look like "Dirac points" are not stable on their own, one merely needs to add a "mass-term" $m \sigma_3$ with $m \in \RM\setminus \{0\}$ to open a gap in the spectrum. However, they do become stable if one enforces onto $H$ an additional chiral symmetry, {\it i.e.} one requires it to anti-commute with a self-adjoint unitary matrix $J$ which here one can choose to be $\sigma_3$. One can then associate to isolated band-crossings at energy zero a charge in odd $K$-theory $K_1(\SM^1)$ and has similar stability properties under deformations as in the odd-dimensional case as long as the chiral symmetry is preserved.

\section{Complex Clifford algebras}
\label{sec:clifford}

This section briefly recalls the definition and some well-known properties of the complex Clifford algebras. The complex Clifford algebra $\CM_d$ of $d$ generators is the universal $\CM$-algebra generated by selfadjoint elements $\Gamma_1, \ldots, \Gamma_d$ satisfying the anti-commutation relation $\Gamma_i \Gamma_j + \Gamma_j \Gamma_i = 2 \,\one\,\delta_{i,j}$. They admit $2^{\lfloor \frac{d}{2} \rfloor}$-dimensional irreducible representations, which are unique up to unitary equivalence for even $d$, but for odd $d$ there are two equivalence classes distinguished by the sign in
$$
\Gamma_1 \cdots \Gamma_d \;=\; \pm\, \imath^{\frac{d+1}{2}}
\;.
$$
Indeed, the product is only fixed to commute with all generators, thus a scalar by Schur's lemma, and to be a self-adjoint unitary or an anti-self-adjoint unitary depending on the parity of $\frac{d-1}{2}$. Irreducible matrix representations can be generated iteratively by starting with the Pauli matrices
$$
\sigma_1 \,=\, 
\begin{pmatrix} 0 & 1\\ 1 & 0
\end{pmatrix}\;, 
\quad 
\sigma_2 \,=\, 
\begin{pmatrix}
0 & \imath\\ -\imath & 0
\end{pmatrix}
\;, 
\quad 
\sigma_3 \,=\, 
\begin{pmatrix}
1 & 0 \\ 0 & -1
\end{pmatrix}
\;,
$$
of which $(\sigma_1,\sigma_2)$ generate an irreducible representation of $\CM_2$ and $(\sigma_1,\sigma_2,\sigma_3)$ one of two irreducible representations of $\CM_3$. The other can be obtained {\it e.g.} by permuting two of those generators. Given a representation $(\sigma_1,\ldots,\sigma_{d-2})$ of $\CM_{d-2}$ with $d>1$ odd, one obtains generators of $\CM_{d-1}$ and $\CM_{d}$ by setting 
$$
\Gamma_i 
\;=\; 
\begin{pmatrix}
0 & \sigma_i \\ \sigma_i & 0
\end{pmatrix}
\quad 
\forall\; \,
i=1,\ldots,d-2\,, 
\qquad 
\Gamma_{d-1} 
\;=\; 
\begin{pmatrix}
0 & \imath \\ -\imath & 0
\end{pmatrix}
\;,
\quad 
\Gamma_{d} 
\;=\; 
\begin{pmatrix}
\one & 0 \\ 0 & -\one
\end{pmatrix}
\;.
$$
With those conventions one has for odd $d$ the irreducible representation with $\Gamma_1 \cdots \Gamma_d = \imath^{\frac{d-1}{2}}$, which will be called the left-handed convention ({\it e.g.} as in \cite{SS1}). Note that in Proposition~\ref{prop-Generators} both constructions use irreducible representations of an odd-dimensional Clifford algebra, but we do not specify the handedness. 

\begin{lemma}
It is sufficient to prove {\rm Proposition~\ref{prop-Generators}} under the additional assumption that the irreducible representation of $\CM_d$ respectively $\CM_{d+1}$ is precisely the one which was iteratively constructed above.
\end{lemma}

\noindent {\bf Proof.} If the representation is not left-handed one first applies the homeomorphism 
$$
\SM^d \,\to\, \SM^d\; , \quad (x_1,\ldots ,x_d)\,\mapsto\, (-x_1,x_2\ldots ,x_d)
\;,
$$ 
and replaces $\Gamma_1 \mapsto -\Gamma_1$ after which the generator $Q^W_d$ respectively $U_d^D$ again takes the same form, but is constructed from a left-handed representation, which is therefore unitarily equivalent to the one given above.
\hfill $\Box$

\vspace{.2cm}

A different handedness of the representation corresponds to a relative minus sign of the generators, hence one needs to be careful with these conventions if it is relevant for the outcome of a computation.  

\section{The connecting maps of $K$-theory}
\label{sec:connectingmap}

Let $\Aa$ be a C$^*$-algebra. Recall ({\it e.g.} \cite{GS,LS1}) that the $K_0$-group of $\Aa$ can be defined as the quotient 
$$
K_0(\Aa)\;=\;
\Vv_0(\Aa)\slash\sim_0
$$ 
where
\begin{equation}
\label{eq-V0def}
\Vv_0(\Aa)
\;=\;
\left\{
V\in \cup_{n\geq 1}M_{2n}(\Aa^+)\;:\;
V\,=\,V^*\;\mbox{\rm invertible}\;,
\;\;
s(V)\sim_0 E_{2n}
\right\}
\;,
\end{equation}
where $M_{2n}(\Aa^+)$ denotes the $2n\times 2n$ matrices over unitalization $\Aa^+$, and $s(V)\sim_0 E_{2n}$ requires the scalar part $s(V)$ of $V$ to be homotopic to $E_{2n}=E_2^{\oplus^n}$ with $E_{2}=\binom{\one\;\;\;\;0\;}{0\;\;-\one}$ within the space of scalar matrices  of adequate size $2n$. On $\Vv_0(\Aa)$ an equivalence relation $\sim_0$ is defined by homotopy (w.r.t. the C$^*$-norm topology) within the self-adjoint invertibles of a given fixed matrix algebra $M_{2n}(\Aa^+)$ and alternatively the requirement 
\begin{equation}
\label{eq-equirel}
V\;\sim_0\;
\begin{pmatrix}
V & 0 \\
0 & E_2  
\end{pmatrix}
\;\in\;M_{2(n+1)}(\Aa^+)\;,
\qquad
V\,\in\,M_{2n}(\Aa^+)
\;.
\end{equation}
Then $K_0(\Aa)$ becomes an abelian group with neutral element $0=[E_2]_0$ via
\begin{equation}
\label{eq-semigroup}
[V]_0\;+\;[V']_0\;=\;\left[
\begin{pmatrix}
V & 0 \\ 0 & V'
\end{pmatrix}
\right]_0
\;.
\end{equation}
The inverse in $K_0(\Aa)$ is given by $-[V]_0=[-V]_0$ and furthermore  $0=[E_{2n}]_0$ for all $n\geq 1$.

\vspace{.2cm}

The standard way to introduce the group $K_1(\Aa)$ is to set
\begin{equation}
\label{eq-K1def}
\Vv_1(\Aa)
\;=\;
\left\{
A\in \cup_{n\geq 1}M_{n}(\Aa^+)\;:\;
A\;\mbox{\rm invertible}
\right\}
\;,
\end{equation}
and to define an equivalence relation $\sim_1$ by homotopy and $[A]_1=[\binom{A\;0}{0\;\one}]_1$. Then $K_1(\Aa) =\Vv_1(\Aa)/\sim_1$ with addition again defined by $[A]+[A']=[A\oplus A']$. If $\Aa$ is unital, one can work with $M_n(\Aa)$ instead of $M_n(\Aa^+)$ in $\Vv_1(\Aa)$, without changing the definition of $K_1(\Aa)$. {Let us note that one can use the unitary operators in \eqref{eq-K1def} instead of the invertibles, simply because they are a deformation retract obtained by the polar decomposition. In the following,  elements of $K_1(\Aa)$ will be represented as unitary operators.}

\vspace{.2cm}

$K$-theory connects the $K$-groups of a given short exact sequence 
\begin{equation}
\label{eq-ShortExact}
0 \;\rightarrow \;\Kk \;\rightarrow\; \Aa\;\overset{\pi}{\rightarrow} \;\Qq \;\rightarrow \;0 
\end{equation}
of C$^*$-algebras in an associated $6$-term exact sequence of abelian groups \cite{RLL,WO}:
\begin{diagram}
& K_0(\Kk)  &  & \rTo{i_*} & & K_0(\Aa) & & \rTo{\pi_*} & & K_0(\Qq) &
\\
&   \uTo_{\Ind}  &   & & & & &   & & \dTo_{\Exp} &  \\
&    K_1(\Qq) &     &  \lTo{\pi_*}   &           & K_1(\Aa) & & \lTo{i_*} & & K_1(\Kk) &
\end{diagram}
Let us write out the two connecting maps, the index map $\Ind$ and the exponential map $\Exp$, in a convenient form. 

\begin{proposi}[{\cite[Exercise 8.D]{WO}} or \cite{LS1}]
\label{prop-IndexMap}
Let the contraction $B\in M_n(\Aa^+)$ be a lift of a unitary $U\in M_n(\Qq^+)$, namely $\pi^+(B)=U$ where $\pi^+:\Aa^+\to\Qq^+ $ is the natural extension of $\pi$ in \eqref{eq-ShortExact}. Then
\begin{equation} 
\label{classic_boundary_of_unitary}
\Ind( [U]_1)
\;=\;[V]_0
\;,
\qquad
V\;=\;
\begin{pmatrix}
2BB^{*}-\one & 2B\sqrt{\one-B^{*}B}\\
2B^{*}\sqrt{\one-BB^{*}} & \one-2B^{*}B
\end{pmatrix}
\;.
\end{equation}
\end{proposi}

\noindent {\bf Proof } (included from \cite{LS1} for the convenience of the reader). First of all, let us note that indeed $V\in\Kk^+$ is a self-adjoint unitary $V\in\Kk^+$ with $s(V)\sim_0 E_{2n}$ because $\pi^+(2BB^*-\one)=\one$ and $\pi^+(\one-2B^*B)=-\one$, and $B^{*}\sqrt{\one-BB^{*}}=\sqrt{\one-B^*B}\,B^*$. The definition of $\Ind$ as given in \cite{RLL} uses a lift $W\in\Alge^+$ of $\diag(U,U^*)$ {that is of norm $\|W\|\leq 1$,} and is 
\begin{equation}
\label{eq-IndIntermed}
\Ind( [U]_1)
\;=\;
\varphi_0
\left(
\left[W
\begin{pmatrix}
\one & 0 \\ 0 & 0
\end{pmatrix}
W^*\right]
\;-\;
\left[
\begin{pmatrix}
\one & 0 \\ 0 & 0
\end{pmatrix}
\right]
\right)
\;,
\end{equation}
where $\varphi_0$ is the map defined in \cite[Proposition~10]{GS} identifying the standard projection picture of $K_0(\Aa)$ from \cite{RLL} to \eqref{eq-V0def}. Due to \eqref{eq-equirel}, this eliminates the second summand in \eqref{eq-IndIntermed} and leads to 
$$
\Ind( [U]_1)
\;=\;
\left[
2\,W
\begin{pmatrix}
\one & 0 \\ 0 & 0
\end{pmatrix}
W^*
\,-\,\begin{pmatrix}
\one & 0 \\ 0 & \one
\end{pmatrix}
\right]
\;.
$$
Choosing
$$
W
\;=\;
\begin{pmatrix}
B &  -\sqrt{\one-BB^*} \\ \sqrt{\one-B^*B}  & B^*
\end{pmatrix}
\;,
$$
now concludes the proof.
\hfill $\Box$

\begin{proposi}
\label{prop-ExpMap}
Let the contraction $B=B^*\in M_n(\Aa^+)$ be a selfadjoint lift of a selfajoint unitary $V\in M_n(\Qq^+)$, namely $\pi^+(B)=V$ where $\pi^+:\Aa^+\to\Qq^+ $ is the natural extension of $\pi$ in \eqref{eq-ShortExact}. Then
\begin{equation} 
\label{eq-ExpMap}
\Exp([V]_0)
\;=\;[B\sqrt{\one-B^2}\;+\;\imath\,(\one-2B^2)]_1
\;.
\end{equation}
\end{proposi}

\noindent {\bf Proof.} The standard picture of the exponential map is as follows \cite{RLL}: let  $P=\frac{1}{2}(V+\one)$; then a selfadjoint lift of $P$ is $\frac{1}{2}(B+\one)$ so that
$$
\Exp([P]_0)\;=\;
\left[
\exp\big({-}2\pi\imath(\tfrac{1}{2}(B+\one))\big)
\right]_1
\;=\;
[-\,\cos(\pi B)\,{+}\,\imath\,\sin(\pi B)]_1
\;.
$$
The crucial point here is that $\cos(\pi B)-\one\in\Kk$ and $\sin(\pi B)\in\Kk$, and that the appearing operator is a unitary operator. However, one can deform these two functions within the invertible operators to the functions $2B^2-\one$ and $B\sqrt{\one-B^2}$. One such homotopy is
$$
t\in[0,1]\;\mapsto\;A_t
\;=\;
\big(-t\cos(\pi B)+(1-t^2)(2B^2-\one)\big)
\;+\;\imath\big(t\sin(\pi B)+(1-t^2)B\sqrt{\one-B^2}\big)
\;.
$$
By spectral calculus, one readily checks that this operator $A_t$ remains invertible for all $t\in[0,1]$. Now $A_1=(2B^2-\one)+\imath B\sqrt{\one-B^2}$. Deforming the global phase by $\pi$ leads to  \eqref{eq-ExpMap}.
\hfill $\Box$

\section{Six-term exact sequence of the sphere}
\label{sec:sixterm}


The $6$-term exact sequence associated to \eqref{eq-SphereExact} is
\begin{diagram}
& K_0(C_0(\DM^{d+1}))  &  & \rTo{i_*} & & K_0(C(\overline{\DM^{d+1}})) & & \rTo{\pi_*} & & K_0(C(\SM^d)) &
\\
&   \uTo_{\Ind}  &   & & & & &   & & \dTo_{\Exp} &  \\
&    K_1(C(\SM^d)) &     &  \lTo{\pi_*}   &           & K_1(C(\overline{\DM^{d+1}})) & & \lTo{i_*} & & K_1(C_0(\DM^{d+1})) &
\end{diagram}
Now $\overline{\DM^{d+1}}$ is a contractible compact space, which implies that the topological $K$-theory are  $K^0(\overline{\DM^{d+1}})=\ZM$ and $K^{-1}(\overline{\DM^{d+1}})=0$ ({\it e.g.} Corollary 2.1.8 and 2.3.10 in \cite{Par}). This implies that $K_0(C(\overline{\DM^{d+1}}))=\ZM$ and $K_1(C(\overline{\DM^{d+1}}))=0$. Furthermore,  $C_0(\DM^{d+1})\cong C_0(\RM^{d+1})$ since the two underlying topological spaces are homeomorphic. 

\vspace{.2cm}

Via the inverse of the stereographic projection $\pi^{-1}:\RM^{d}\mapsto \SM^{d}\setminus \{e_{d+1}\}$ the sphere without north pole $e_{d+1}=(0_d,1)$ is identified with the one-point compactification of $\RM^{d}$, hence $C_0(\RM^{d})^+ \cong  C(\SM^{d})$. Concretely, let us fix the bijections  $y\in\DM^{d+1}\mapsto \sigma(y)=z\in\RM^{d+1}\mapsto \pi^{-1}(z)=x\in\SM^{d+1}\setminus\{e_{d+1}\}$ where $\sigma$ is scaling function $\sigma$ and $\pi^{-1}$ the inverse of the stereographic projection $\pi$ given by
$$
z\,=\,\sigma(y)\,=\,
\frac{y}{\sqrt{1-\|y\|^2}}\;,
\qquad
x\,=\,\pi^{-1}(z)
\,=\,\Big(\frac{2z}{1+\|z\|^2},\frac{\|z\|^2-1}{1+\|z\|^2}\Big)
\;.
$$
Then $\sigma_*$ identifies $K_i(C_0(\RM^{d}))\cong K_i(C_0(\DM^{d+1}))$ and $\pi^{-1}_*$ provides an embedding  $K_i(C_0(\RM^{d+1}))\subset K_i(C(\SM^{d+1}))$. In particular, one can write 
$$
K_i(C_0(\DM^{d+1}))\,=\,K_i(C_0(\RM^{d+1}))\,=\,\widetilde{K}_i(C(\SM^{d+1}))
\;,
$$ 
where the reduced $K$-group is again defined by dividing out the subgroup generated by the constant functions in $C(\SM^{d+1})$. One can now read off the $K$-groups \eqref{eq-KSphere} from the six-term exact sequence, which is for odd $d$ is given by
\begin{diagram}
& \ZM &  & \rTo{i_*} & & \ZM & & \rTo{\pi_*} & & K_0(C(\SM^d)) &
\\
&   \uTo_{\Ind}  &   & & & & &   & & \dTo_{\Exp} &  \\
&   K_1(C(\SM^d)) &     &  \lTo{\pi_*}   &           & 0 & & \lTo{i_*} & & 0 &
\end{diagram}
while for even $d$ 
\begin{diagram}
& 0 &  & \rTo{i_*} & & \ZM & & \rTo{\pi_*} & & K_0(C(\SM^d)) &
\\
&   \uTo_{\Ind}  &   & & & & &   & & \dTo_{\Exp} &  \\
&   K_1(C(\SM^d)) &     &  \lTo{\pi_*}   &           & 0 & & \lTo{i_*} & & \ZM &
\end{diagram}
The middle $\ZM$ in the upper rows is generated by the unit element of $C(\overline{\DM}^{d+1})$, hence $i_*$ is the zero-map and $\pi_*$ injective. The diagrams therefore split into two group extension problems, one for each of the two abelian groups $K_0(C(\SM^d))$ and $K_1(C(\SM^d))$, with the unique solution \eqref{eq-KSphere}.

\vspace{.2cm}

The first diagram also shows that the index map induces an isomorphism:
\begin{equation}
\label{eq-IndSphere}
\Ind:K_1(C(\SM^d)) \to K_0(C_0(\DM^{d+1}))=\widetilde{K}_0(C(\SM^{d+1}))
\;,
\qquad
d\;\mbox{\rm odd}\;.
\end{equation}
In the second diagram for even $d$, the trivial part of $K_0(C(\SM^d))=\ZM\oplus\ZM$ is the image of $\pi_*:K_0(C(\overline{\DM^{d+1}}))=\ZM\to K_0(C(\SM^d))$. The second, non-trivial part, is then mapped bijectively onto $K_1(C_0(\DM^{d+1}))=K_1(C(\SM^{d+1}))=\ZM$ by
\begin{equation}
\label{eq-ExpSphere}
\Exp:\widetilde{K}_0(C(\SM^d)) \to K_1(C_0(\DM^{d+1}))=K_1(C(\SM^{d+1}))
\;,
\qquad
d\;\mbox{\rm even}\;.
\end{equation}
For the proof of Proposition~\ref{prop-Generators} it thus only remains to show that maps in \eqref{eq-IndSphere} and \eqref{eq-ExpSphere} indeed map the Dirac Hamiltonian onto the Weyl Hamiltonian and vice versa:
\begin{equation}
\label{eq-ExpInd}
\Ind ([U^D_d]_1)\;=\;[Q^W_{d+1}]_0\;,
\qquad
\Exp([Q^W_d]_0)\;=\;[U^D_{d+1}]_1
\;.
\end{equation}

\section{Index map of odd-dimensional spheres}

Let $d$ be odd and write $U^D_d(y)=\sum_{j=1}^dy_j\,\gamma_j+\imath\, y_{d+1}$ where $y=(y_1,\ldots,y_{d+1})\in\SM^d\subset\RM^{d+1}$ with $\|y\|=1$ and $\gamma_1,\ldots,\gamma_d$ is a given irreducible, self-adjoint representation of the Clifford algebra $\CM_d$. To compute $\Ind[U^D_d]_1$ by \eqref{classic_boundary_of_unitary}, let us note that $y\in\DM^{d+1}\mapsto \sum_{j=1}^dy_j\,\gamma_j+\imath\, y_{d+1}$ is indeed a contraction lift of $U^D_d$. Hence
$$
\Ind([U^D_d]_1)
\,=\,
\left[
\begin{pmatrix}
2\|y\|^2-1 & \!\!\! 2(1-\|y\|^2)^{\frac{1}{2}}\big(\sum_{j=1}^dy_j\,\gamma_j\;+\;\imath\, y_{d+1}\big)
\\
2\big(\sum_{i=1}^dy_j\,\gamma_j\;-\;\imath\, y_{d+1}\big)(1-\|y\|^2)^{\frac{1}{2}} & -(2\|y\|^2-1)
\end{pmatrix}
\right]_0
,
$$
where the r.h.s. is understood as a function on $\DM^{d+1}$, namely $\|y\|^2=\sum_{j=1}^{d+1}y_j^2<1$ and not equal to $1$. Let the matrix on the r.h.s. be denoted by $G_{d+1}(y)$. It is indeed $E_2$ with terms vanishing on $\partial\DM^{d+1}$. Hence $\Ind[U^D_d]_1\in K_0(C_0(\DM^{d+1}))$ as needed.  One has generators $\Gamma_1,\ldots ,\Gamma_{d+2}$ of an irreducible representation of $\CM_{d+2}$ by setting (as in Section~\ref{sec:clifford})
$$
\Gamma_j\;=\;\gamma_j\otimes\sigma_1
\;,
\qquad
\Gamma_{d+1}
\;=\;
\one\otimes\sigma_2\;,
\qquad
\Gamma_{d+2}
\;=\;
\one\otimes\sigma_3\;,
$$
for $j=1,\ldots,d$ and then finds
$$
G_{d+1}(y)
\;=\;
2(1-\|y\|^2)^{\frac{1}{2}}
\Big(
\sum_{j=1}^{d+1}y_j\,\Gamma_j
\Big)
\;+\;
(2\|y\|^2-1)\Gamma_{d+2}
\;.
$$
If one maps $x\in \SM^{d+1}$ to $\RM^{d+1}$ using the stereographic projection and then rescales to the unit disc via $\sigma: \RM^{d+1}\to \DM^{d+1}$ one ends up with
$$
x
\;=\;\pi^{-1}\circ\sigma(y)
\;=\;
\big(2y\sqrt{1-\|y\|^2},2\|y\|^2-1\big)
\;.
$$
Thus 
$$
G_{d+1}(\sigma^{-1}\circ\pi(x))
\;=\;
\sum_{j=1}^{d+2}x_j\,\Gamma_j
\;,
$$
which is hence indeed the Weyl Hamiltonian restricted to the sphere  $\SM^{d+1}$, which proves the first equality in \eqref{eq-ExpInd}.

\section{Exponential map of even dimensional spheres}

Let $d$ be even and $Q^W_d(y)=\sum_{j=1}^{d+1}y_j\Gamma_j$ where $y=(y_1,\ldots,y_{d+1})\in\SM^d$, namely $\|y\|=1$. 
It clearly has selfadjoint contraction lift given by $y\in\DM^{d+1}\mapsto B(y)=\sum_{j=1}^{d+1}y_j\,\Gamma_j$. Now using \eqref{eq-ExpMap}, one obtains
\begin{equation}
\label{eq-ExpExplicit}
\Exp([Q^W_d]_0)
\;=\;\big[B(y)\sqrt{\one-\|y\|^2}\;+\;\imath\,(2\|y\|^2-\one)\big]_1
\;,
\end{equation}
where $\|y\|^2=\sum_{j=1}^{d+1}(y_j)^2=B^2(y)$. Now note that 
$$
x_j\;=\;y_j\sqrt{\one-\|y\|^2}\;\;\;\;\mbox{\rm for }j=1,\ldots,d+1\;,
\qquad
x_{d+2}\;=\;2\|y\|^2-1\;,
$$
are actually the coordinate of a new odd-dimensional $(d+1)$-sphere and that the r.h.s. of \eqref{eq-ExpExplicit} is exactly the associated unitary $U^D_{d+1}$, namely the second equality in \eqref{eq-ExpInd}.


\section{Relation to the Fredholm picture}
\label{sec:fredholm}

In this section, the generators of $K_i(C_0(\RM^{d+1}))$ given in Proposition~\ref{prop-Generators} are used to determine the unbounded generators of the isomorphic group $KK_i(\CM, C_0(\RM^{d+1}))$. 

\vspace{.2cm}

Let us start out be explaining this isomorphism which is merely the so-called Fredholm picture of $K$-theory  (see \cite[Section 17.5.1]{Bla} or \cite{PS2}). It states that any class in $K_0(\Aa)$ can be represented by a bounded adjointable operator $T$ on a standard Hilbert $\Aa$-module $\Hh_\Aa=\Aa \otimes \Hh$, with separable but possibly finite-dimensional Hilbert space $\Hh_\Aa$, satisfying the Fredholm conditions
\begin{equation}
\label{fredholm_cond}
\one\,-\,T^*T\, ,\; \one\,-\,TT^* \;\in \;\KM(\Hh_\Aa)
\;.
\end{equation}
For $K_1(\Aa)$, on top of the condition \eqref{fredholm_cond} one requires the representatives $T$ to be self-adjoint. To be precise, $T$ (together with the Hilbert module it acts on) defines a class in $KK_i(\CM, \Aa)$ which is isomorphic to $K_i(\Aa)$ via the maps
$$
\Ii_0\;:\; [T]_0 \,\mapsto \,\Ind([T+\KM(\Hh_\Aa)]_1)
\;, 
\qquad \Ii_1\;:\; [T]_1 \,\mapsto \,\Exp([T+\KM(\Hh_\Aa)]_0)
\;,
$$
where the index and exponential map are w.r.t. the extension
$$
0\; \rightarrow \;\KM(\Hh_\Aa) \;\rightarrow\; \BM(\Hh_\Aa) \;\rightarrow\; \BM(\Hh_\Aa)/\KM(\Hh_\Aa) \;\rightarrow\; 0
\;.
$$
This is well-defined since $T+\KM(\Hh_\Aa)$ is a (self-adjoint) unitary and $K_i(\Hh_\Aa)\cong K_i(\Aa)$ due to stability. Let us briefly recall why this is an isomorphism. One can assume that $\Hh$ is some fixed infinite-dimensional Hilbert space (if $T$ acts on a different Hilbert module $\Kk_\Aa$ one can always extend $T$ to $T \oplus \one_{\Hh_\Aa}$ and use the absorption theorem $\Kk_\Aa\oplus \Hh_\Aa\simeq \Hh_\Aa$). This gives an isomorphism of $KK_i(\CM, \Aa)$ with $K_{1-i}(\BM(\Hh_\Aa)/\KM(\Hh_\Aa))$ and since the $K$-groups of $\BM(\Hh_\Aa)$ are trivial, exponential and index map are both isomorphisms.

\vspace{.2cm}

In the so-called unbounded picture one instead uses unbounded regular operators $T$ with compact resolvent, {\it i.e.} $T$ and $T^*$ are densely defined,  $\one+T^*T$ has dense range and the Fredholm condition \eqref{fredholm_cond} is replaced by 
\begin{equation}
\label{eq:fredholm_cond_unbounded}
(\one+T^*T)^{-1}\,, \;(\one+TT^* )^{-1}\; \in\; \KM(\Hh_\Aa)
\;.
\end{equation}
This is accomplished simply by replacing $T$ with its bounded transform $F(T):= T (\one+T^*T)^{-\frac{1}{2}}$, which is then a bounded operator that satisfies \eqref{fredholm_cond}.

\vspace{.2cm}

Let us now consider the case $\Aa=C_0(\RM^{d+1})\otimes \CM_{d+1}$ where $\CM_{d+1}$ is the complex Clifford algebra with $d+1$ generators. Associated is the Hilbert module $\Hh_\Aa=C_0(\RM^{d+1})\otimes \CM^{N}$ with $N$ being the dimension of the Clifford representation. Now introduce, with the same notations as in Proposition~\ref{prop-Generators}, the unbounded functions
$$
x\, \in\, \RM^{d+1} \;\mapsto \;H^W_d(x)\,=\,\sum_{j=1}^{d+1} x_j \Gamma_j \in M_N(\CM)
\;,
$$
for even $d$ and
$$
x\, \in\, \RM^{d+1} \;\mapsto \;A^D_d \,=\, \sum_{j=1}^{d} x_j \Gamma_j\, +\, \imath x_{d+1}
$$
for odd $d$. These operators define unbounded multipliers on the Hilbert module $C_0(\RM^{d+1})\otimes \CM^{N}$. They have compact resolvents since both square to $\lvert x\rvert^2$ and $x\in \RM^{d+1} \mapsto (1+\lvert x\rvert^2)^{-1}$ is a $C_0$-function and the compact operators on $C_0(\RM^{d+1})\otimes \CM^{N}$ are $M_N(C_0(\RM^{d+1})$.  Since $H^W_d$ is self-adjoint, one has classes  $[H^W_d]_1 \in KK_1(\CM, C_0(\RM^{d+1}))$ for odd $d$, respectively $[A^D_d]_0 \in KK_0(\CM, C_0(\RM^{d+1}))$ for even $d$. The following result is also folklore, and again we could not {locate} a detailed proof.

\begin{proposi}
\label{prop-KKGen}
$[H^W_d]_1$ respectively $[A^D_d]_0$ are unbounded normalized generators of the groups $KK_i(\CM, C_0(\RM^{d+1}))$ for $d+1=i\;\mbox{\rm mod}\; 2$.
\end{proposi}

\noindent {\bf Proof.}
Let us now argue that using the identifications $K_i(C_0(\DM^{d+1}))=K_i(C_0(\RM^{d+1}))\subset K_i(C(\SM^{d+1}))$ as above, one has
\begin{equation}
\label{eq:rel_bott}
\Ii_1([H^W_d]_1) \;=\;\Exp([Q^W_{d}]_0)\;,
\qquad 
\Ii_0([A^D_d]_0)  \;=\;\Ind([U^D_{d}]_1)\;,
\end{equation}
where $\Exp$ and $\Ind$ are the connecting maps associated to \eqref{eq-SphereExact}. Then the claim follows from the equalities $\Exp([Q^W_{d}]_0)=[U_{d+1}^D]_1$ and $\Ind([U^D_{d}]_1)=[Q_{d+1}^W]_0 $ given in \eqref{eq-ExpInd}.

\vspace{.2cm}

For the proof of \eqref{eq:rel_bott}, let $\rho: C_0(\DM^{d+1})\to C_0(\RM^{d+1})$ denote the isomorphism induced by the homeo\-morphism $\sigma: \DM^{d+1}\to \RM^{d+1}$ from Section~\ref{sec:sixterm}. Like any homomorphism, it extends uniquely to the multiplier algebras and thus to a homomorphism $\rho: C(\overline{\DM^{d+1}})\to M(C_0(\RM^{d+1}))=C_b(\RM^{d+1})$ (the image of the homomorphism $\rho$ can be characterized as the $C^*$-algebra $C_{\mathrm{vg}}(\RM^d)$ generated by functions whose gradients vanish at infinity). One therefore has a commutative diagram with exact rows
\begin{diagram}
	0 & \rTo{} & & C_{0}(\DM^{d+1}) &  & \rTo{} & & C(\overline{\DM^{d+1}}) & & \rTo{} & & C(\SM^{d}) & & \rTo{} & 0
	\\ & & &  \dTo_{\rho}  &   & & & \dTo_{\rho} & &   & & \dTo_{\hat{\rho}} &  \\
	0 &	\rTo &
	& C_0(\RM^{d+1}) &  & \rTo{} & & M(C_0(\RM^{d+1})) & & \rTo{} & & M(C_0(\RM^{d+1}))/C_0(\RM^{d+1}) & & \rTo & 0
\end{diagram}
where $\hat{\rho}(f)$ is defined by choosing any function $\hat{f}\in C_b(\RM^{d+1})$ whose radial limits exist and are equal to $f\in C(\SM^{d})$, then setting $\hat{\rho}(f) = \hat{f} + C_0(\RM^{d+1})$. In particular, $\hat{\rho}(Q^W_d)= F(H^W_d)+ M_N(C_0(\RM^{d+1}))$, since the bounded transform of $H^W_d$ is the function
$$
x \,\in \,\RM^{d+1}\; \mapsto\; (F(H^W_d))(x) \,= \,\sum_{j=1}^d \frac{x_j}{\sqrt{1+x^2}}\, \Gamma_j
\;.
$$
Similarly one has $\hat{\rho}(U^D_d)=F(A^D_d)+M_N(C_0(\RM^{d+1}))$. Naturalness of the $K$-theoretic connecting maps implies that one can use either of the exact sequences to evaluate the exponential map, namely \cite[Propositions 9.1.5 and 12.2.1]{RLL}
$$
\Exp\big(\big[F(H^W_d)+ M_N(C_0(\RM^{d+1}))\big]_0\big) 
\;=\; 
\rho_*\big( \Exp([Q^W_d]_0)\big)
\;, 
$$
and 
$$
\Ind\big(\big[F(A^D_d)+ M_N(C_0(\RM^{d+1}))\big]_1\big) \;=\; \rho_*\big(\Ind([U^D_d]_1)\big)
\,,
$$
which are exactly the equalities between the left-most and right-most terms in \eqref{eq:rel_bott} since $\rho_*$ is implicit under the used identifications.
\hfill $\Box$

\vspace{.3cm}

\noindent {\bf Acknowledgements:} This work was partially supported by the DFG grants SCHU 1358/8-1 and STO 1454/1-1. 

\noindent {\bf Competing interests statement:} The authors have no competing interests. 



\begin{thebibliography}{99}
\bibliographystyle{unsrt}

\bibitem{BvE} P.~F.~Baum, E.~van~Erp, {\sl $K$-homology and Fredholm operators I: Dirac operators}, J. Geometry Physics {\bf 134}, 101-118 (2018).

\bibitem{Bla} B.~Blackadar, {\sl $K$-theory for operator algebras}, (Cambridge University Press, Cambridge, 1998).

\bibitem{BL} J.~L.~Boersema, T.~A.~Loring, {\sl $K$-theory for real $C^*$-algebras via unitary elements with symmetries},
New York J. Math. {\bf 22}, 1139-1220 (2016).

\bibitem{CSB} A.~L.~Carey, H.~Schulz-Baldes, {\sl Spectral flow of monopole insertion in topological insulators},
Commun. Math. Phys. {\bf 370}, 895-923 (2019).

\bibitem{Get} E.~Getzler, {\sl The odd Chern character in cyclic homology and spectral flow}, Topology {\bf 32},  489-507 (1993).

\bibitem{GS} J.~Grossmann, H. Schulz-Baldes, {\sl Index pairings in presence of symmetries with applications to topological insulators}, Commun. Math. Phys. {\bf 343}, 477-513 (2016).

\bibitem{Horava} P. H\v{o}rava, {\sl Stability of Fermi Surfaces and $K$-Theory}, Phys. Rev. Lett. {\bf 95}, 016405 (2005).

\bibitem{JM} C.~M.~Joseph, R.~Meyer, {\sl Geometric construction of classes in van Daele $K$-theory}, J. Math. Phys. {\bf 64}, 053504 (2023).

\bibitem{K1} M.~Karoubi, {\sl Lectures on K-Theory}, in {\sl Cohomology of Groups and Algebraic K-theory}, Advanced Lectures in Mathematics Vol. 12 (International Press of Boston, Somerville, MA, 2010).

\bibitem{LS1} T.~Loring, H.~Schulz-Baldes, {\sl Finite volume calculation of $K$-theory invariants}, New York J. Math. {\bf 23}, 1111-1140 (2017).

\bibitem{Par} E.~Park, {\sl Complex topological $K$-theory}, (Cambridge Univ. Press, Cambridge, 
2008).

\bibitem{PS} E.~Prodan, H.~Schulz-Baldes, {\sl  Bulk and boundary invariants for complex topological insulators: From $K$-theory to physics}, (Springer Int. Pub., Cham, Szwitzerland, 2016).

\bibitem{PS2} E.~Prodan, H.~Schulz-Baldes, {\sl  Generalized Connes-Chern characters in $KK$-theory with an application to weak invariants of topological insulators}, Rev. Math. Phys. {\bf 28}, 1650024 (2016).

\bibitem{RLL} M.~Rordam, F.~Larsen, N.~Laustsen, {\sl An Introduction to K-theory for C$^*$-algebras}, (Cambridge University Press, Cambridge, 2000).

\bibitem{SS1} H.~Schulz-Baldes, T.~Stoiber, {\sl Semiclassical  localization for semimetals and Callias operators}, {J. Math. Phys. {\bf 64}, 081901 (2023).}

\bibitem{WO} N.~E.~Wegge-Olsen, {\sl K-theory and C$^*$-algebras}, (Oxford
Univ. Press, Oxford, 1993).


\end{thebibliography}
\end{document}